\documentclass[twocolumn,english,aps,prl]{revtex4-1}
\usepackage[T1]{fontenc}
\usepackage[latin9]{inputenc}
\usepackage{float}
\usepackage{amsmath}
\usepackage{amssymb}
\usepackage{graphicx}
\usepackage{esint}

\makeatletter
\@ifundefined{textcolor}{}
{%
 \definecolor{BLACK}{gray}{0}
 \definecolor{WHITE}{gray}{1}
 \definecolor{RED}{rgb}{1,0,0}
 \definecolor{GREEN}{rgb}{0,1,0}
 \definecolor{BLUE}{rgb}{0,0,1}
 \definecolor{CYAN}{cmyk}{1,0,0,0}
 \definecolor{MAGENTA}{cmyk}{0,1,0,0}
 \definecolor{YELLOW}{cmyk}{0,0,1,0}
}

\usepackage{babel}

\makeatother

\usepackage{babel}
\begin{document}
\global\long\global\long\def\bra#1{\left\langle #1\right|}

\global\long\global\long\def\ket#1{\left|#1\right\rangle }

\global\long\global\long\def\bk#1#2#3{\bra{#1}#2\ket{#3}}

\global\long\global\long\def\ora#1{\overrightarrow{#1}}

\global\long\global\long\def\vev#1{\left\langle #1\right\rangle }

\title{Spin Transfer Torque and Electric Current in Helical Edge States in Quantum Spin Hall Devices}

\author{Qinglei Meng, Smitha Vishveshwara, and Taylor L. Hughes} 
\affiliation{Department of Physics, University of Illinois, 1110 West Green St, Urbana IL 61801}

\begin{abstract}
We study the dynamics of a quantum spin Hall edge coupled to a magnet with its own 
dynamics. Using spin transfer torque principles, we analyze the interplay between
 spin currents in the edge state and dynamics of the axis of the magnet, and draw parallels with circuit analogies. 
As a highlighting feature, we show that while coupling to a magnet typically renders the edge state insulating
 by opening a gap, in the presence of a small potential bias,  spin-transfer torque can restore perfect 
conductance by transferring angular momentum to the magnet. In the presence of  interactions within the edge 
state, we employ a Luttinger liquid treatment to show that the edge, when subject to a small voltage bias, tends to 
form a unique dynamic rotating spin wave state that naturally couples into the dynamics of the magnet. 
We briefly discuss realistic physical parameters and constraints for observing this interplay between quantum 
spin Hall and spin-transfer torque physics.
\end{abstract}

\maketitle
The study of symmetry protected topological insulators has led to a series of remarkable theoretical and experimental discoveries\cite{hasan2010,hasan2011,maciejko2011}. The initial prediction of the time-reversal invariant quantum spin Hall (QSH) insulator\cite{kane2005a,kane2005b} was soon realized in HgTe/CdTe quantum wells\cite{bernevig2006c,konig2007} and, more recently, in InAs/GaSb quantum wells\cite{liu2008,knez2011}. One of the most exciting features of the quantum spin Hall insulator is the presence of robust, gapless edge states with counter propagating modes with opposite spin-polarization. The edge states form a so-called helical liquid which is a new class of 1D liquids that is perturbatively stable as long as time-reversal symmetry is preserved\cite{wu2006,xu2006}.  As predicted, experiments show that each edge, when biased, exhibits a quantized two-terminal, longitudinal conductivity of $e^2/h,$ even in the presence of disorder, as long as time-reversal symmetry is not broken\cite{bernevig2006c,konig2007,konig2008,roth2009,maciejko2011}.

Some of the most interesting observable predictions concerning helical modes involve proximity coupling of the edge states to magnetic and/or superconducting layers that act to de-stabilize the edge and open a gap\cite{qi2008,qi2008b,fu2008}. For example, in the non-interacting limit, the helical liquid is simply a 1D Dirac fermion and it is well-known that the domain-walls of mass-inducing perturbations in this system lead to topological bound states\cite{jackiw1976,su1979}. These bound states can be the source of fractional charges, for the case of a magnetic domain wall\cite{jackiw1976,su1979,qi2008}, or Majorana zero modes in a superconductor-magnet interface\cite{fu2008}. Our focus is on the coupling of magnetic perturbations to the helical liquid.
Several works have discussed the coupling of dynamical magnets to the QSH edge states leading to various effects like adiabatic charge pumping\cite{thouless1983,qi2008}, a spin-battery effect\cite{mahfouzi2010}, and many related effects in 3D time-reversal invariant topological insulators\cite{garate2010,yokoyama2010,burkov2010,mahfouzi2012,ma2012,chen2013}.

 In this article we consider the transport properties of a helical liquid in proximity to a dynamical ferromagnetic island and show that this coupled magnet-QSH edge system can exhibit a rich range of behavior due to spin-transfer torque physics. In the generic case, the magnetic island opens a gap and acts as a barrier for the helical liquid via the Zeeman coupling (if its magnetization has a component perpendicular to the spin-polarization of the helical liquid). Essentially the island provides a local mechanism for up-spin right-moving modes to backscatter to down-spin left-moving modes. This is a process that is usually strongly suppressed as it would  require the electrons on one edge to scatter across the gapped interior of the sample to the opposite edge. Thus, in light of the gap formation, we would naively expect that if the edge is coupled to the magnetic island and then voltage biased, then as long as the voltage does not exceed the magnet-induced gap then there will be no longitudinal conductance on this edge. During this process we would expect charge to build up near the island until the potential counter-acts the applied voltage to reach a steady-state, i.e., we should have capacitor-like physics. In this article we show that, unexpectedly, once the magnet is dynamically influenced by a spin-transfer torque applied by the scattering edge states, then the edge can begin conducting current even at zero-temperature when the applied voltage is smaller than the magnet-induced gap. The resulting electrical behavior is then inductor-like.

\begin{figure}
\begin{centering}
\includegraphics[scale=0.235]{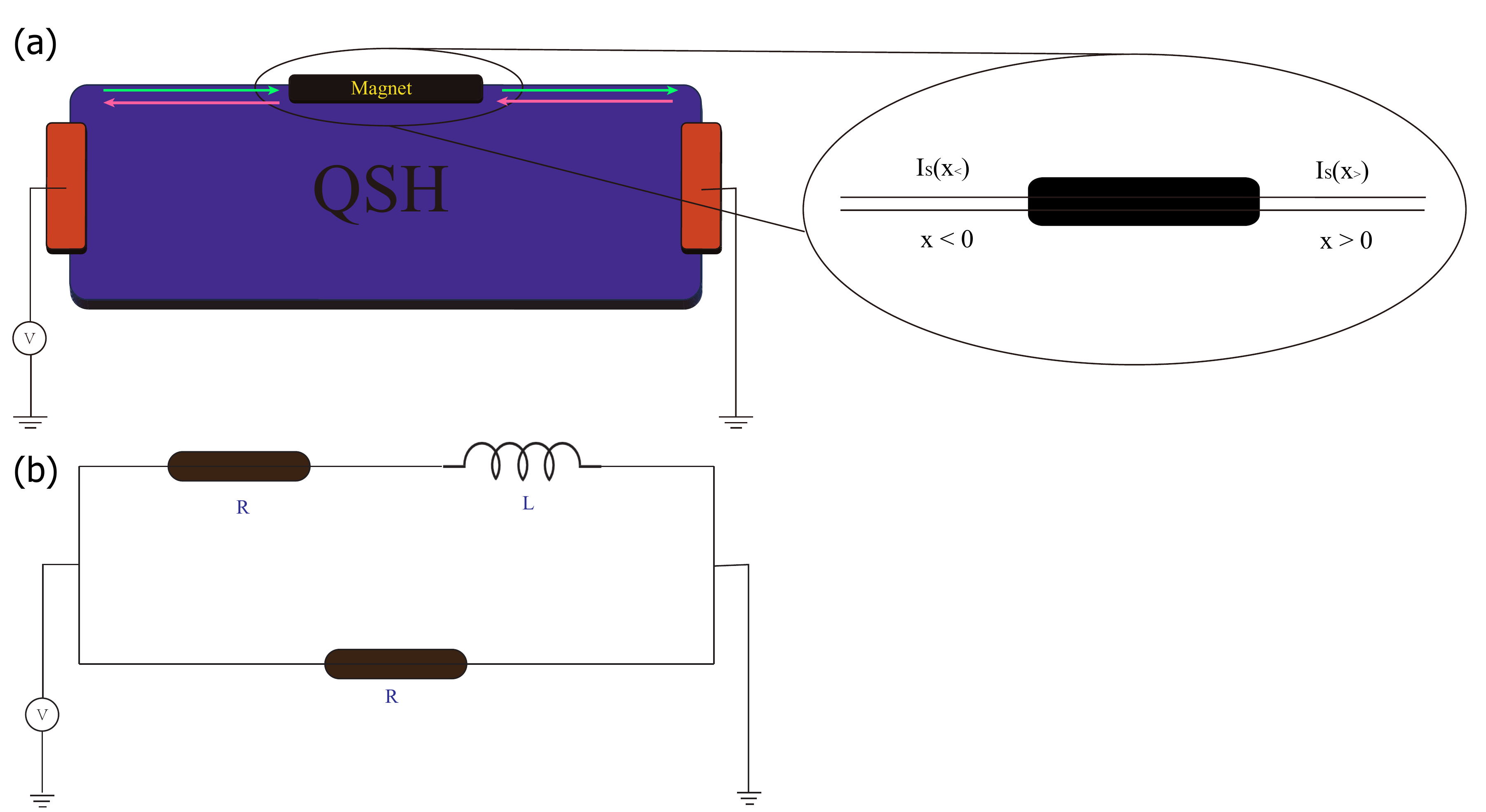}
\par\end{centering}
\caption{Quantum spin Hall insulator system coupled to a magnetic island on a single edge in the presence of a finite bias voltage. (a) We schematically illustrate the basic setup and zoom in the region of the island. (b) Circuit analogy for the Hall bar and magnet. Resistors represent the quantized $h/e^2$ resistance of each edge and the inductor represents the effect of the magnetic island. 
\label{fig:QSH-voltage}}
\end{figure}

The basic idea is as follows. A standard unbiased QSH bar carries
gapless chiral edge currents of opposite spin (say polarized along $\hat{z}$
) traveling in opposite directions, thus carrying zero charge 
current but two 	quantized units of spin current, if we ignore spin relaxation for now. The presence of a magnet,
as in Fig. \ref{fig:QSH-voltage}, couples the left and right movers and gaps these modes if the magnetization is not entirely along $\hat{z}.$ The gap renders the QSH edge a charge insulator, in that there is no initial
charge current for a bias voltage with associated energy less than
that of the magnet-induced gap. However, because of the spin-momentum locking, taking into account the spin degree of freedom yields more
complex behavior. In the presence of such a bias the edge initially carries
excess spin current on one side of the magnet. This imbalance of spin-current on the sides of the magnet provides
a spin torque that results in the transfer of angular momentum and subsequent bias-controlled dynamics of the magnet. Again, because of the spin-momentum locking, the induced dynamics of the magnet in turn affects
the QSH dynamics. For instance, in spite of the charge gap exceeding the
applied voltage, the magnet induces a charge current
in the edge as it rotates\cite{qi2008} due to the spin-transfer torque. Hence, we show
 that the magnet can act as an inductive circuit element instead of a capacitive element.

 In what follows, we model the QSH-magnet coupled system and explore its dynamics
employing spin-transfer torque methods.  We analyze the approach to
steady state, the nature thereof and characteristic relaxation times,
and draw parallels with electrical circuit analogies. Applicable to experimental
realizations, we estimate the effect from typical parameters of the QSH in HgTe/CdTe quantum wells and with the magnetic system
of $K_{2}CuF_{4}$\cite{yamada1972}. We then study the interplay
between magnetization dynamics and bias voltage in the presence of interactions
in the QSH edge states. Previously, within a Luttinger liquid framework,
we have shown an instability towards an unusual spin-density wave ordering\cite{meng2012a};
here we find that the bias voltage endows this textured phase with
unique dynamics.

\begin{figure}
\centering{}\includegraphics[scale=0.14]{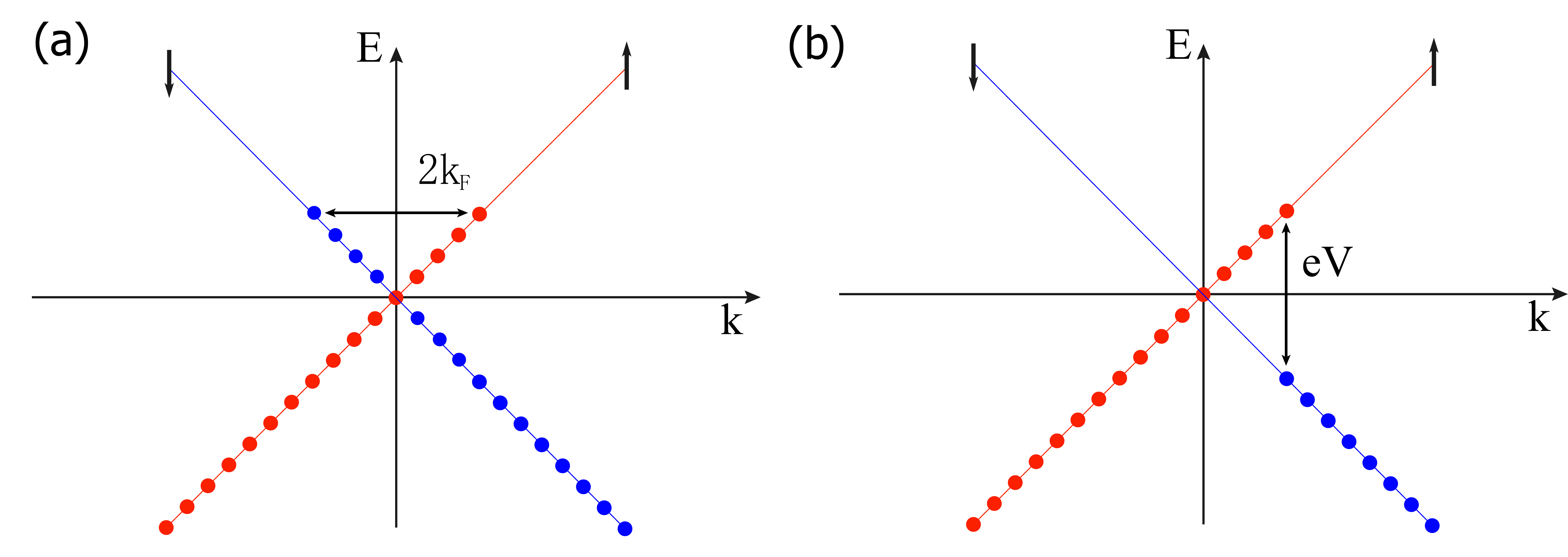}\caption{(a) Spectrum of the free helical liquid at finite chemical potential. With repulsive interactions present, system forms a gapped spin-density wave order parameter with wave vector $2k_F$, which nests the Fermi-points. (b) Spectrum of a free current-carrying helical liquid in the presence of a finite bias voltage. Alternatively, when coupled to a magnet,in the rotating frame of the magnet,  the helical edges show a relative chemical potential shift.  The magnetic order parameter that can effectively gap the associated Fermi-points has to thus connect states at different energies, exhibiting finite frequency dynamics.\label{fig:Specturm-of-QSH}}
\end{figure}

Beginning with the free helical liquid, let us consider the QSH edge description which has the associated Hamiltonian\cite{wu2006,xu2006}
\begin{eqnarray}
H_{0} & = & \int dx\hbar v\left[\psi_{L\uparrow}^{\dagger}(x)(i\partial_{x})\psi_{L\uparrow}(x)
  -  \psi_{R\downarrow}^{\dagger}(x)(i\partial_{x})\psi_{R\downarrow}(x)\right]\nonumber.\\\label{eq:QSH-1}
\end{eqnarray}
As shown in Fig. \ref{fig:Specturm-of-QSH}, these correspond to linearly dispersing  edge states moving along the $x$-direction with speed $v,$ where the operator $\psi_{R\uparrow(L\downarrow)}$ annihilates an
electron moving to the right(left) with up(down) spin. The proximity coupling
between the magnet with magnetization $\ora M=(M_{x},M_{y},M_{z})$
and the QSH edge is given by the usual Zeeman coupling
\begin{equation}
H_{M}=-\mu_{0}\mu_{B}\ora M\cdot\ora{\sigma}\label{eq:QSHmag}
\end{equation}
where $\mu_{0}$ is the the vacuum permeability, $\mu_{B}$ is
the Bohr magneton and the Pauli matrices $\ora{\sigma}=(\sigma_{x},\sigma_{y},\sigma_{z})$
act on the space $\psi=[\psi_{R\uparrow}(x),\psi_{L\downarrow}(x)]^{T}$. In the region near the magnet, the QSH edge spectrum is effectively $\sqrt{(vp-\mu_{0}\mu_{B}M_{z})^{2}+(\mu_{0}\mu_{B})^{2}(M_{x}^{2}+M_{y}^{2})}$
which has an excitation gap induced by the magnet with magnitude $\Delta=2\mu_{0}\mu_{B}(M_{x}^{2}+M_{y}^{2})^{1/2}.$


Let us now consider the effect of a voltage bias $V$, which, for instance, we apply at the lead
on the left in Fig.~\ref{fig:QSH-voltage}. Initially the spin currents
in the left and right  regions of the magnet are different, giving
rise to a spin current imbalance $\Delta\ora I_{S}=[\ora I_{S}(x_{<})-\ora I_{S}(x_{>})]$,
where $\ora I_{S}(x)=\frac{\hbar}{2}\psi^{\dagger}\frac{1}{2}(v\sigma_{z}\ora{\sigma}+\ora{\sigma}v\sigma_{z})\psi$,
and $x_{<}(x_{>})$ is on the left (right) of the magnet as indicated
in  Fig. \ref{fig:QSH-voltage}a. Because of the spin-momentum locking of the helical liquid, the spin
current imbalance generically depends on the rotation frequency of the magnet. For simplicity, let us consider the case when the magnet rotates in-plane
at a frequency ($\frac{\Omega_M}{2\pi}$). We can transform the full edge Hamiltonian $H=H_{0}+H_{M}$
to the rotating frame via the transformation $H'=UHU^{\dagger}-iU\partial U^{\dagger}$,
where $U=e^{i\frac{\Omega_M t}{2}\sigma_{z}}$\cite{mahfouzi2010}. In the new basis, the Hamiltonian
takes the resultant form 
\begin{equation}H_{{\rm{rot}}}=\left[\begin{array}{cc}
\hbar(-iv\partial_{x}-\frac{\Omega_M}{2}) & -\mu_{0}\mu_{B}M_{s}\\
-\mu_{0}\mu_{B}M_{s} & \hbar(iv\partial_{x}+\frac{\Omega_M}{2})
\end{array}\right],
\end{equation} where there is a rotation-induced voltage shift of $\hbar \Omega_M/e$ which is opposite for each spin component (see Fig. \ref{fig:Specturm-of-QSH}b). After imposing
the appropriate boundary conditions and matching $\psi$ fields at the interfaces between the unperturbed helical liquid and the magnet, we
find the initial spin current imbalance  
\begin{equation}
\Delta\ora I_{S}(t=0)=\frac{eV-\hbar \Omega_M}{2\pi}\hat{z}.\label{eq:imbalanced spin current}
\end{equation}
Here have assumed that the length of magnet, $L_{M}$,
is long enough , $L_{M}\gg\frac{\hbar v}{\mu_{0}\mu_{B}M_{s}}$, that
a low-energy electron incident on the magnet barrier is completely
reflected; accounting for tunneling requires a simple modification. 



The spin-current imbalance applies a torque on the magnet and we can appeal to spin transfer torque (STT) physics to analyze the coupled
dynamics between QSH edge currents and the magnet. Applying the well-established
STT formalism\cite{Slonczewski1996,berger1996}, the dynamics is
described by the Landau-Lifshitz equation
\begin{equation}
\gamma^{-1}\partial_{t}\ora M=-D\ora M\times M_{z}\hat{z}+\frac{1}{V_{M}}\hat{M}\times(\Delta\ora I_{S}\times\hat{M})\label{eq:spin transfer torque}
\end{equation}
where $\gamma$ is the gyromagnetic ratio of the magnet, $V_{M}$
is the volume of the magnet, and $\hat{M}$ is the unit vector directed
along the magnetization $\ora M$. The first term on
the right-hand side accounts for the easy-plane anisotropy energy $\frac{1}{2}DM_{z}^{2}V_{M}$ of the magnet
\cite{Zhang1998}. The source of the anisotropy can be either intrinsic, as for an easy-plane magnet, or induced
by the coupling to the edge states itself, though the latter effect is weak compared to usual magnetic energy scales. Thus we would generally
desire the intrinsic anisotropy to be large to observe interesting dynamics
since the edge coupling is usually small. The second term on the right-hand
side accounts for the torque due to the spin current imbalance $\Delta\ora I_{S}$ derived above.
Since the magnitude of the magnetization is effectively fixed, the spin torque
along the direction of the magnetization has no effect; only the transverse
part of this imbalanced spin current exerts the torque on the magnetization.
We will see that the effect of this term is to drive the edge from an insulating state to
a conducting state. 


In general, the dynamics derived from substituting the spin imbalance expression of Eq. (\ref{eq:imbalanced spin current})
into the dynamical equation of motion Eq. (\ref{eq:spin transfer torque}) has no simple
solution. However, in most  of the physical cases of interest we can make the approximation that
the magnet always stays in-plane, i.e. $M_{z}\ll M_{S}$, where $M_{S}$
is the magnitude of the spontaneous magnetization. This condition
holds for small enough bias voltages, i.e., bias voltages that are small compared
to the magnet-induced gap, as will be justified in the proposed experimental setup to follow.
With this approximation, we obtain the simple solution:
\begin{eqnarray}
\Delta\ora I_{S} & =&\frac{eV}{2\pi}e^{-\frac{\gamma^{2}D\hbar}{2\pi V_{M}}t}\hat{z}\nonumber\\
I_{C} & =&\frac{e^{2}V}{h}(1-e^{-\frac{\gamma^{2}D\hbar}{2\pi V_{M}}t})\nonumber\\
 M_{x}+iM_{y}&=&M_{S}e^{i\int_0^t \Omega_Mdt'}\nonumber\\
 M_{z} & =&\frac{2\pi}{e\gamma D}I_{C}\label{eq:LLsol}
\end{eqnarray}\noindent where $I_{C}=e\psi^{\dagger}v\sigma_{z}\psi$ is the charge current
on the edge. Thus, we can immediately see that the dynamics involves a characteristic relaxation
time $\tau=\frac{2\pi V_{M}}{\gamma^{2}D\hbar}$. The smaller the
magnet and larger the anisotropy, the faster the relaxation. 

We can simply illustrate the consequences of the dynamics. The 
 STT on the magnet due to the spin current imbalance ($\Delta\ora I_{S}$)
decays to zero, while the in-plane magnetization begins to rotate; the rotation frequency increases to the constant value $\frac{eV}{h}$ . Interestingly, in spite
of the magnet-induced gap, the charge current ramps up to its quantized saturation value of $\frac{e^{2}}{h}V$,
rendering the magnetic barrier transparent to charge. The
spin transfer torque provides a magnetization along the $z$ direction,
which reaches a new equilibrium value $\frac{eV}{\gamma D\hbar}$.
In fact, this $z$ direction magnetization acts as an effective magnetic
field causing the in-plane magnetization to precess. The charge current that flows here is essentially due to the same charge-pumping mechanism reported in Ref. \onlinecite{qi2008} for a rotating magnet. However, for our case the magnetization dynamics and the rotation frequency are intrinsically controlled by the applied bias voltage. 

An even simpler picture for understanding the dynamics involves representing the geometry in Fig. \ref{fig:QSH-voltage}a
as an effective electrical circuit analog shown in  Fig. \ref{fig:QSH-voltage}b.
The upper and lower QSH edges in Fig. \ref{fig:QSH-voltage} each provide
a resistance of $R=\frac{h}{e^{2}}$. What our dynamical solution has shown is that the coupling of the edge states to the magnet
can effectively be represented by an inductor with inductance $L=\tau R$.
Hence, for this set up, charge is only transported through the lower edge initially,
which yields an effective conductance of $e^{2}/h$. As with a real inductor,
which stores energy in an induced field, here the energy is stored
in the form of the anisotropy energy of the easy-plane magnet. Over time,
the inductive component becomes transparent, allowing current to pass
through. In the final steady state, the upper and lower edges both
conduct perfectly and the conductance of the system rises and saturates
to its quantized value of $2e^{2}/h.$

Let us briefly consider a physical magnetic system, for which we focus on $K_{2}CuF_{4}$,
known for its large easy-plane anisotropy\cite{yamada1972,Hirakawa1979}. This
material possesses a spontaneous magnetization of $\mu_{0}M_{s}=0.124\text{T},\ $
a gyromagnetic ratio $\gamma=-2\times10^{11}\text{s}^{-1}\text{T }^{-1}$,
and an out-of-plane anisotropy field $B_{A}=0.280\text{T}$, i.e.
$D=\frac{B_{A}}{M_{s}}=2.26\mu_{0}$. For a typical magnet of volume
$V_{M}=10^4\text{nm}\times10^2\text{nm}\times10^2\text{nm}$ these
parameters provide a relaxation time estimate of $\tau=10^{-1}\text{s}.$
In order to be consistent with our approximation that $M_{z}\ll M_{S}$ ,we require
an applied voltage to be smaller than $1\text{mV }$. This constraint
confines the rotational frequency of the magnet ($\Omega_M=\frac{eV}{h}$) and
the associated radiation to lie in the microwave range which indicates that microwave cavity resonator experiments may be useful for the observation of this effect. 

While we have so far presented a clean, optimistic description of the effect,  it must be mentioned that in addition to the primary contribution
to the dynamics stemming from spin transfer torque, one also expects
two sources of dissipation: (i) Gilbert damping of the magnetization dynamics and (ii) spin-relaxation of the helical liquid due to spin-orbit scattering. Gilbert damping contributes an additional
term $\frac{\alpha}{M_{s}\gamma}\ora M\times\frac{d\ora M}{dt}$ to
the right-hand side of Eq. (\ref{eq:spin transfer torque}) , where
$\alpha$ is the damping constant. As shown in Appendix \ref{app:damping} we find that the damping provides
an additional channel for relaxation, modifying the relaxation rate
in Eq. (\ref{eq:LLsol}) to $\tau^{-1}=\gamma^{2}D(\frac{\hbar}{2\pi V_{M}}+\alpha\frac{M_{S}}{|\gamma|})$.
More importantly, it also changes the precession frequency to $\Omega_M=eV/(\hbar+\alpha2\pi V_{M}M_{S}/|\gamma|)$. The effects of spin-orbit scattering will cause the spin carried by the helical liquid to relax as the charge current is carried from the leads to the magnetic island. This will reduce the amount of spin-current imbalance by a geometry and impurity-dependent factor $\zeta$ and subsequently the precession frequency will be reduced by the same factor. Both of these effects alter $\Omega_M,$ and since the charge current is simply $e\Omega_M,$ these two sources of dissipation will reduce the saturation conductance of the magnet-coupled edge from its quantized value. Notably, experiments indicate that spin-orbit scattering effects do not dominate the spin physics in HgTe/CdTe quantum wells\cite{brune2012}, however the Gilbert damping of the magnet will surely reduce the effective edge conductance, though hopefully not below an observable value.

So far we have neglected interactions within the QSH edges; we now
examine the stability of the magnetization dynamics presented above in the presence of interactions. First we will
 consider the possibility of new interaction-driven phenomena, initially analyzing the QSH system in and 
of itself without the coupling to the magnet. As done previously\cite{wu2006,xu2006,hou2009,meng2012a,meng2012b},
the interacting helical liquid can be explored within a Luttinger liquid framework
through the bosonization of the fermion fields. We can use the boson fields $\phi$
and $\theta$ and the correspondence $\psi_{R\uparrow}(x)\sim e^{-i(\phi(x)-\theta(x))},\ \psi_{L\downarrow}(x)\sim e^{i(\phi(x)+\theta(x))}$ to bosonize the helical liquid.
The Hamiltonian in Eq. (\ref{eq:QSH-1}), along with interactions,
can be bosonized to yield the Luttinger liquid Hamiltonian\cite{Giamarchi}
\begin{equation}
H=\frac{1}{2\pi}\int dx\left[uK(\nabla\theta)^{2}+\frac{u}{K}(\nabla\phi)^{2}+2\mu\nabla\phi\right],
\label{eq:LL}
\end{equation}
where $u=v((1+\frac{g_{4}}{2\pi v})^{2}-(\frac{g_{2}}{2\pi v})^{2})^{1/2}$
is the renormalized velocity,  $K=(\frac{1+\frac{g_{4}}{2\pi v}-\frac{g_{2}}{2\pi v}}{1+\frac{g_{4}}{2\pi v}+\frac{g_{2}}{2\pi v}})^{1/2}\ $
is the Luttinger parameter, and the $g_2, g_4$ represent the standard interaction coupling constants\cite{Giamarchi}. Values of $K<(>)1\ $represent repulsive
(attractive) interactions, and here we only consider repulsive interactions. We have also included a chemical potential
($\mu$) term to account for the edge Fermi-level not lying
exactly at the Dirac point, a condition that leads to interesting physics in the presence of interactions (see Fig. \ref{fig:Specturm-of-QSH}a for an illustration in the free case). For repulsive interactions, the system is unstable to spontaneously breaking time-reversal symmetry and generating in-plane ferromagnetic order which opens a gap at the Fermi-energy when $\mu=0.$ If the chemical potential is not exactly
tuned to be at the Dirac point, the system instead exhibits a spatial oscillation
of the in-plane magnetic order, forming spin density wave (SDW) \cite{Giamarchi,meng2012a}.

For the remainder of the calculations it is convenient
to transform into the Lagrangian formulation, yielding the Lagrangian associated with Eq.~\ref{eq:LL}
\begin{equation}
L_{\gamma}=\frac{1}{2\pi uK}((\partial_{t}\phi)^{2}-u^{2}(\nabla\phi)^{2}-2\mu uK\nabla\phi)+\partial_{t}(\gamma\phi). \label{eq:Lagrangian}
\end{equation}
Here the last term corresponds to a total derivative, which
does not affect the classical equations of motion, but allows us to add a non-vanishing charge current. The parameter
$\gamma$ represents an additional freedom that should be fixed
by a physical quantity, which we choose to be the particle current operator
$j$ given by
\[
j=\vev{\frac{\partial_{t}\phi}{\pi}},
\]
thus fixing the choice: $\gamma=-j/Ku$.

The effect of repulsive interactions on the helical liquid can be seen by evaluating the appropriate susceptibilities.
 The primary quantity of interest is the susceptibility of the operator $O_{+}(x,t)\equiv\psi_{R\uparrow}^{\dagger}(x,t)\psi_{L\downarrow}(x,t)$ which is
related to the in-plane magnetization of the edge state via $m_{+}(x,t)\equiv m_{x}(x,t)+im_{y}(x,t)=2\mu_{B}\langle O_{+}(x,t)\rangle$. To evaluate the spin susceptibility associated with the in-plane magnetization,
$\chi_{m}(x,t)=-i\hbar\theta(t)\langle [O_{+}(x,t), O_{+}^{\dagger}(0,0)]\rangle$,
it is easiest to first shift the $\phi$ field in the Lagrangian in Eq.
(\ref{eq:Lagrangian}) as $\tilde{\phi}(x,t)=\phi(x,t)+\mu Kx/u+\pi uK\gamma t$,
and then employ standard Luttinger liquid techniques. As a function of
temperature $T$ we find that the Fourier-transformed susceptibility in momentum and
frequency space diverges as 
\[
\chi_{m}(\omega=-2\pi j, k=-2\tfrac{K}{u}\mu)\sim T^{2K-2}
\]\noindent near $(\omega_c,k_c)=(-2\pi j, -2K\mu/u).$

The divergence of the spin susceptibility is indicative of an intrinsic instability
towards a magnetically ordered phase in the presence of repulsive
interactions. As we discussed in previous work~\cite{meng2012a}, the
finite momentum at which the spin susceptibility diverges indicates that for $\mu\neq 0$ SDW order is preferred in which the
in-plane magnetization spatially rotates over a length scale $\sim\frac{\pi u}{K\mu}$.
A new effect is that, in the presence of an injected current, the susceptibility diverges at finite-frequency, i.e.,  the SDW order rotates
at the frequency $2\pi j$ as a function of time. Thus, the edge can be carrying current and in a gapped, intrinsically-magnetized state if the SDW order rotates as a function of time. 



We can heuristically illustrate why the time oscillation of the SDW occurs by resorting to the free-fermion description
($K=1$) where the current  $j$ induced by a bias voltage $V$ can be determined by the filling of the single-particle energy spectrum as shown in
Fig. \ref{fig:Specturm-of-QSH}b. In the presence of the repulsive
interaction term $H_{int}=\psi_{R\uparrow}^{\dagger}\psi_{R\uparrow}\psi_{L\downarrow}^{\dagger}\psi_{L\downarrow}=O_{+}(x,t)O_{+}^{\dagger}(x,t)$
the system will try to develop in-plane magnetic order $\langle O_{+}(x,t)\rangle$,
in order to induce a mass term $\psi_{R\uparrow}^{\dagger}\psi_{L\downarrow}\langle O_{+}^{\dagger}(x,t)\rangle$,
that will open up a gap and lower the energy of the system. Notice that
the most efficient way to lower the energy is to open up the gap at
the Fermi points. When the current  vanishes this implies that SDW order will form with a wave-vector that nests the two degenerate Fermi-points (see Fig. \ref{fig:Specturm-of-QSH}a). However, when there is finite current in this system,
then, effectively, the two Fermi points are not at the same energy. In order to couple
these two Fermi points that lie at different energies, the SDW has to have a time dependent part
$\langle O_{+}(x,t)\rangle\sim e^{ieVt/\hbar}$ which is exactly why we observe
a divergent spin susceptibility at finite frequency.



Finally, we revisit the coupling to the external magnet in the presence of interactions.
Since the external magnetic island has been assumed to be uniform, we
expect that to achieve the strongest coupling between QSH edge (with SDW order) and the external magnet,
the length of the magnet should be smaller than the SDW wavelength $\sim\frac{\pi u}{K\mu}.$
 The presence of a magnet, as in the non-interacting case, opens up
a gap in the helical liquid. This is easy to see in the Luttinger liquid formalism, where the coupling between the
edge state and external magnet in Eq.(\ref{eq:QSHmag}) has the Sine-Gordon
form $\cos(2\phi-\theta_{H})$, where $\theta_{H}$ is the angle
of the in-plane magnetization. This coupling is relevant in the renormalization
group sense, and hence locks the phase $2\phi=\theta_{H}$ at low temperature.
In previous work, we have analyzed the static effect of external magnets on the helical liquid at finite $\mu$~\cite{meng2012a}. For the dynamic
situation we are considering in this work, one can derive the particle current
as $j=\vev{\frac{\partial_{t}\phi}{\pi}}=\frac{\partial_{t}\theta_{H}}{2\pi}$
which is simply the adiabatic charge pumping on the QSH edge as derived in Ref. \cite{goldstone1981,thouless1983, qi2008}, but now including interactions. 

If the magnet is not initially rotating then, just as in
the non-interacting case, we expect an initial spin current
imbalance across the magnet when a voltage is applied. We can calculate this spin current imbalance, which, due to spin-momentum locking is proportional  
to the charge density difference across the the magnet, $\Delta I_S^z = \frac{\hbar}{2}v(\rho(x_<) - \rho(x_>)) = \frac{K}{2\pi}\frac{v}{u}eV $. Thus, even with interactions there is an initial spin-current imbalance which will apply a STT to the magnet. While a full analysis of the spin-transfer torque in the presence of interactions is beyond
the scope of this work, we expect that just as with the non-interacting case, the excess spin current, now accompanied by an in-plane magnetization rotation of the edge, transfers angular momentum to the magnetic region. Once again, as with the non-interacting case, in steady state, a charge current will flow as the magnet evolves to a steady-state of rotation at a rate proportional to the applied voltage.

\textit{Applications -} The unique combination of QSH physics and
spin transfer torque gives rise to new ways of probing and manipulating
the QSH edge, particularly by exploiting well-characterized magnetic
materials and their information storage and access properties. \textit{i)
Microwave resonator - } We saw above that an excess QSH spin current
produced by a voltage bias $V$ induces the magnet to precess at a
frequency $eV/h$. This precession would result in microwave radiation
 of about $24\text{GHz}$ for typical bias voltages of order $0.1\text{meV}$.
In principle, one can envision putting an array of QSH-coupled magnets
in a microwave resonator to generate a voltage-tunable microwave laser. \textit{ii)
Spin polarization detector- }Thus far, we have assumed that the QSH
spin axis coincides with that of the easy plane of the spin magnet.
In principle, the two need not be aligned, effectively giving the
excess QSH spin current components in the $xy$-plane and in turn
affecting the dynamics of the magnet. Analyzing this dynamics would
provide information on spin polarization in the QSH system.
 iii)\textit{AC QSH circuit}
- Information on the QSH edges can also be obtained by charge current
measurements from the perspective of the circuit analogy of Fig. \ref{fig:QSH-voltage}.
The circuit description can be taken further by including a conventional capacitance element
to produce oscillatory charge and spin currents. 
In conclusion, here we have presented an initial glimpse of the rich physics that can
emerge through the interplay of QSH edge and spin-transfer torque physics.

\begin{acknowledgments}
We are grateful to E. Johnston-Halperin for insightful comments. For support, we acknowledge the U.S. Department of Energy, Division of Materials Sciences under Award No. DE-FG02-07ER46453  (T. L. H. and S. V.) and the National Science Foundation under Grant No. DMR-0906521 (Q. M.).
\end{acknowledgments}

\appendix
\section{Gilbert damping}\label{app:damping}

Now we include the Gilbert damping term in the spin transfer torque analysis of the Landau-Lifshitz-Gilbert equation:

\begin{eqnarray}
\gamma^{-1}\partial_{t}\ora M&=&-D\ora M\times M_{z}\hat{z}+\frac{1}{V_{M}}\hat{M}\times(\Delta\ora I_{S}\times\hat{M})\nonumber
\\&+&\frac{\alpha}{M_{s}\gamma}\ora M\times\frac{d\ora M}{dt}\label{eq:damping}
\end{eqnarray} where the last term is due to Gilbert damping.
Now we can write Eq. \ref{eq:damping} in terms of components, and furthermore continue our approximation from the body of the text where we assume $M_{z}\ll M_{S}$ for total in-plane magnetization $M_S.$ Additionally, making an ansatz that $M_{x}=M_{S}\cos \theta(t),\ M_{y}=M_{S}\sin \theta(t),$

\begin{eqnarray*}
\gamma^{-1}\partial_{t}M_{z} & = & \frac{eV-\hbar \dot{\theta}(t)}{2\pi V_{M}}+\frac{\alpha}{\gamma}M_{S}\dot{\theta}(t)\\
\gamma^{-1}\partial_{t}M_{x} & = & -DM_{S}\sin \theta(t)M_{z}-\frac{\alpha}{\gamma}\dot{\theta}(t)\cos \theta(t)M_{z}\\
\gamma^{-1}\partial_{t}M_{y} & = & DM_{S}\cos \theta(t)M_{z}-\frac{\alpha}{\gamma}\dot{\theta}(t)\sin\theta(t)M_{z}.
\end{eqnarray*}

In general the dynamics can be complicated, even after assuming $M_z\ll M_S.$ Let us consider our physical system of interest  $K_{2}CuF_{4}$  where we estimate that $DM_{S}=0.28\text{T }.$ Thus for small voltages
$V<1\text{mV }$, then $DM_{S}>-\frac{\alpha}{\gamma}\Omega_M$, as long
as $\alpha<10^{-1}.$ With this approximation we have
\begin{eqnarray*}
\gamma^{-1}\partial_{t}M_{z} & = & \frac{eV-\hbar \dot{\theta}(t)}{2\pi V_{M}}+\frac{\alpha}{\gamma}M_{S}\dot{\theta}(t)\\
\gamma^{-1}\partial_{t}M_{x} & = & -DM_{S}\sin\theta(t)M_{z}\\
\gamma^{-1}\partial_{t}M_{y} & = & DM_{S}\cos \theta(t)M_{z}.
\end{eqnarray*} From here we see that Gilbert damping will just provide another
channel for the damping of the imbalanced spin current, which will
decrease relaxation time to
\begin{equation}
\tau =  \left[\left(\frac{\hbar}{2\pi V_{M}}+\alpha\frac{M_{S}}{|\gamma|}\right)\gamma^{2}D\right]^{-1}
\end{equation}\noindent and decrease the rotation frequency to
\begin{equation}\Omega_M  = \frac{eV}{\hbar+\alpha2\pi V_{M}M_{S}/\gamma}.
\end{equation}

%

\end{document}